# Experimental and theoretical study of red-shifted solitonic resonant radiation in photonic crystal fibers and generation of radiation seeded Raman solitons

Surajit Bose, Samudra Roy, Rik Chattopadhyay, Mrinmay Pal and Shyamal K. Bhadra, *Member, OSA*

*Abstract*—The red shifted *solitonic* resonant radiation is a fascinating phase matching phenomenon that occurs when an optical pulse, launched in the normal dispersion regime of photonic crystal fiber, radiates across the zero dispersion wavelength. The formation of such phase-matched radiation is independent of the generation of any optical soliton and mainly governed by the leading edge of input pump which forms a shock front. The radiation is generated at the anomalous dispersion regime and found to be confined both in time and frequency domain. We experimentally investigate the formation of such radiations in photonic crystal fibers with detailed theoretical analysis. Our theoretical predictions corroborate well with experimental results. Further we extend our study for long length fiber and investigate the interplay between red-shifted solitonic resonant radiation and intrapulse Raman scattering (IPRS). It is observed that series of radiation-seeded Raman solitons are generated in anomalous dispersion regime.

## I. INTRODUCTION

DISPERSIVE wave (DW) generation is an exciting nonlinear phenomenon that takes place when an optical soliton, propagating in anomalous dispersion (AD) domain, radiates across the zero-dispersion point of optical fiber [1]-[5]. The underlying mechanism is the phase matching resonance between soliton and linear DW where a significant amount of energy is transferred to the DW which grows in normal dispersion (ND) regime and thereby creates blue shifted frequency components [6]. The third and higher order dispersions (HOD) with their numeric signs play dominant role in controlling such radiations [7]. During supercontinuum generation the blue component of the spectra is mainly generated by DWs [8]-[10]. In that process optical pulse is launched in AD regime and soliton dynamics plays the dominant role in generating different frequency components [8]. In spite of the previous extensive studies, *resonant radiation* (RR) is still drawing the attention of the researcher because of its fascinating and versatile characteristic features. Soliton spectral tunneling [11], Raman–induced DW trapping [12], dual radiation in Kagomè lattice [13], soliton self-frequency shift cancellation [14], accelerating rogue waves [15], negative frequency RR [16] are some of the most interesting topics that have been studied in the context of this phase matched radiation. Recently it is found that the DW radiation is not restricted by the AD pumping where the formation of optical soliton is the essential criteria in generating RR [17]-[20]. Indeed this is an important observation in understanding the fundamental physics of RR. Unlike the previous case, we observe that the RR is generated in AD domain, which makes them robust, and is found to be solitonic in nature. Here the word *solitonic* is used because the RR is confined both in frequency and time domain. This is the most fascinating feature of such radiation which characteristically separates them from the conventional DWs where the radiation pulse disperses in time domain. In ND pumping, the self-phase modulation (SPM)-induced frequency chirping and even weak dispersive effect lead to significant pulse shaping. The optical pulse becomes nearly rectangular with relatively sharp leading and trailing edges. The interference of different frequency components propagating at different group velocities leads to *optical wave breaking,* a phenomenon where pulse edges become oscillating [21]. The immediate consequence of optical wave breaking is the generation of dispersive shock wave (DSW) which leads to a phase matched RR under HOD perturbations [19]. The phase matching condition between the pump and radiation wave gives us the expression of the detuned frequency at which radiation takes place.

In this present work we have experimentally studied the generation of phase matched RR in a fabricated photonic crystal fiber (PCF) and tried to investigate its solitonic nature. Femtosecond and picosecond pulses are launched in ND domain close to the zero dispersion wavelength (ZDW) of the PCF and we examine the evolution of red-shifted solitonic RR

January 19, 2015; this work was supported by CSIR-NMITLI , CSIR 12th Plan GLASSFIB project and SRIC- IIT Kharagpur.

Surajit Bose, Rik Chattopadhyay, Mrinmay Pal, Shyamal K. Bhadra are with Fiber Optics and Photonics Laboratory, CSIR-Central Glass and Ceramic Research Institute, 196, Raja S.C. Mullick Road, Kolkata-700 032, India (e-mail:surajit291@gmail.com;chatterjee.rik.84@gmail.com;mpal@cgcri.res.in; skbhadra@cgcri.res.in).

Samudra Roy is with the Department of Physics, Indian Institute of Technology Kharagpur, Kharagpur-721302, India (e-mail: samudraroy@gmail.com).

at AD regime. The nonlinear pulse dynamics differ significantly for femtosecond and picosecond pulses. The Raman scattering is more predominant for shorter pulses and it is interesting to study the interplay between the Raman induced frequency red shift and the phase matched RR which is also generated at higher wavelength side. Our theoretical simulation reveals that the generated RR seeds a series of Raman solitons in AD domain of the fiber. We use generalized nonlinear Schrödinger equation to model the experimental results and try to understand the underlying mechanism of solitonic RR.

## II. EXPERIMENTAL DETAILS

The Schematic of the experimental set up is shown in Fig. 1. We have used two pulse lasers both operating at the same wavelength of 1064 nm but of different pulse widths. One laser source emits a 275 femtosecond pulse at a 50 MHz repetition rate. The maximum average output power is 1 W. The other laser source operates at 5.5 picosecond pulse with 36.7 MHz repetition rate. The maximum average power is 5W. Laser delivery is from a passively mode locked fiber laser with pre-amplifier and high power cladding pumped fiber amplifier with an extra pulse compression unit in the former.

The output of the laser is coupled into the core of the PCF by utilizing a 40x microscope objective lens ($L_1$) with numerical aperture of 0.65 and XYZ-metric (translational) stage. The input end of the nonlinear PCF is properly cleaved and mounted on a three-axis translational stage for perfect alignment so that maximum power is coupled into the fiber. Light emerged from the fiber output is collimated by employing a lens system ($L_2$) and then it is passed through wedged plate beam splitter (BS). The reflected portion is focussed using lens $L_3$ and analysed spectrally by an optical spectrum analyser (OSA) with a measurement range from 600 to 1700 nm. Modal pattern of the transmitted light is recorded using CCD camera. In this work only the fundamental mode of the PCF is excited. Numerous spectral analyses are executed by utilizing different pump power and also by varying the length of the fiber. By fixing the power and the input launch condition, spectra are recorded for different fiber lengths. In Fig. 2 (a) we show the cross section of the suspended core PCF which was used in the experiment. Our target is to pump the PCF at the normal dispersion regime close to ZDW.

The fiber is fabricated by stack and draw process keeping in mind of the proximity of the pulsed laser source of wavelength 1064nm. In order to achieve this, we have drawn silica capillaries of specific dimensions that are manually assembled by stacking the capillaries hexagonally about a central rod made by silica. All are suprasil F-300 grade silica glass. The preform stack is inserted into the glass tube formed by silica, fixed in a furnace of the capillary drawing tower to obtain an intermediate 3mm cane by maintaining slow drawing rate and differential pressure with an optimized temperature. Later fibers of diameter 125 ± 0.5 micron are drawn encapsulating the cane in another tube with online polymer resin coating in another fiber drawing tower. By adjusting the drawing parameters with proper suspension we have fabricated the fiber with desired ZDW. The nonlinearity and the modal confinement of the fiber are enhanced by suspending the core. It is suspended by controlling different drawing parameters like furnace temperature and fiber draw speed. The average air-hole diameter ($d$), pitch ($\Lambda$) and core diameter ($d_c$) are nearly $7.5\,\mu m$, $8.00\,\mu m$ and $6.2\,\mu m$ respectively. The nonlinear coefficient of this fiber is calculated to be $\gamma \approx 7.689 W^{-1} km^{-1}$ at 1064 nm. Dispersion of the fundamental mode (FM) in PCF is calculated by using full vector mode solver software COMSOL Multiphysics based on finite element method (FEM). Maxwell's equations are solved which determines the effective indices ($n_{eff}$) of the guided modes. The group velocity dispersion profile ( $\beta_2(\lambda) = \frac{\lambda^3}{2\pi c^2} \frac{d^2 n_{eff}}{d\lambda^2}$) of the fundamental mode is shown in Fig. 2(b) where it crosses the zero dispersion point at $\lambda_{ZD} \approx 1100$ nm. Fig 2 (c) shows the field distribution of the fundamental mode (FM) captured by CCD camera which ensures that the spectrum generated at the output is for FM only and there is no intermodal coupling. The confinement of the FM at $\lambda_0 = 1064$ nm is also depicted in Fig. 2 (d) calculated using COMSOL Multiphysics.

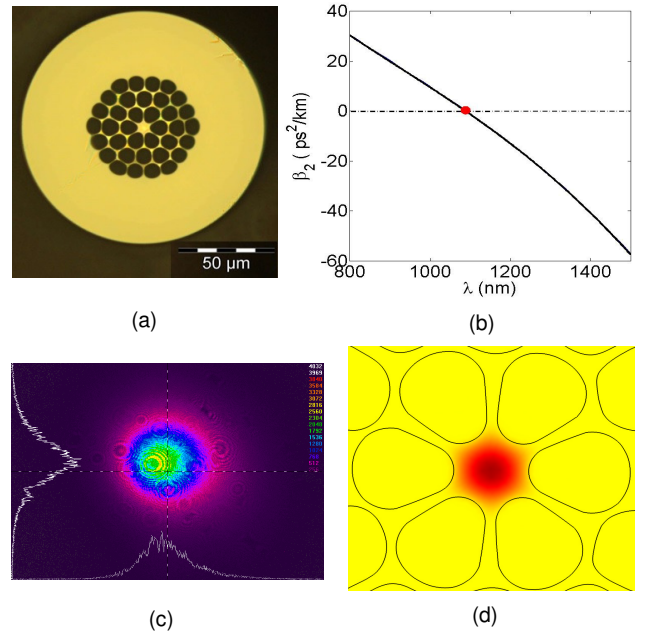

(a) (b)

(c) (d)

Fig. 2: (a) Cross-section of the photonic crystal fiber (PCF) that was employed in the experiment with $d/\Lambda \approx 0.94$. (b) Dispersion profile of the fundamental mode. The red dot indicates the zero dispersion point which is $\lambda_{ZD} \approx 1100$ nm. (c) Distribution of the fundamental mode (FM) recorded experimentally using CCD camera. (d) Distribution of the fundamental mode (FM) at the operating wavelength of $\lambda_0 = 1064$ nm which is obtained numerically using FEM - commercial software COMSOL Multiphysics.

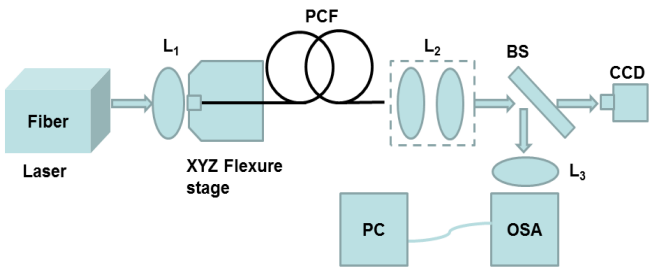

Fig. 1: Schematic diagram of the experimental set-up for generating red shifted resonant radiation

## III. NUMERICAL ANALYSIS

### A. Simulation

In order to capture the complete pulse dynamics inside the fabricated PCF, we employ a generalized nonlinear Schrödinger equation (GNLSE) written in a normalized form as follows [6], [8], [22]:

$$\frac{\partial u}{\partial \xi} = \frac{i}{2} \frac{\partial^2 u}{\partial \tau^2} + \sum_{m \geq 3}^{\infty} i^{m+1} \delta_m \frac{\partial^m u}{\partial \tau^m} + iN^2 \left(1 + i\tau_{sh} \frac{\partial}{\partial \tau}\right) \times \left(u(\xi, \tau) \int_{-\infty}^{\tau} R(\tau - \tau') |u(\xi, \tau')|^2 d\tau'\right) \quad (1)$$

where the field amplitude $u(\xi, \tau)$ is normalized such that $u(0,0) = 1$ and the other dimensionless variables are $\xi = \frac{z}{L_D}$, $\tau = \frac{t - z/v_g}{T_0}$, $N = \sqrt{\gamma P_0 L_D}$ and $\delta_m = \frac{\beta_m}{m! T_0^{m-2} |\beta_2|}$. Here, $N$ defines soliton order, $P_0$ is the peak power of the ultrashort pulse launched into the fiber, $T_0$ is the input pulse width, $\beta_m$ is the $m^{th}$ order dispersion coefficient in real unit i.e. ps$^m$/km, $L_D = T_0^2 / |\beta_2|$ is the dispersion length, $\gamma$ is the nonlinear parameter of the fiber, $\delta_m$ is the $m^{th}$ order dispersion coefficient in dimensionless form and $\tau_{sh} = (\omega_0 T_0)^{-1}$ is the self-steepening parameter at the carrier frequency $\omega_0$ of the pulse. The nonlinear response function of the optical fiber has the form [23],

$$R(\tau) = (1 - f_R)\delta(\tau) + f_R h_R(\tau) \quad (2)$$

where the first and the second terms correspond to the electronic and Raman responses, respectively, with $f_R = 0.245$. As discussed in [23], the Raman response function can be expressed in the following form,

$$h_R(\tau) = (f_a + f_c) h_a(\tau) + f_b h_b(\tau) \quad (3)$$

where the functions $h_a(\tau)$ and $h_b(\tau)$ are defined as,

$$h_a(\tau) = \frac{\tau_1^2 + \tau_2^2}{\tau_1 \tau_2^2} \exp\left(-\frac{\tau}{\tau_2}\right) \sin\left(\frac{\tau}{\tau_1}\right) \quad (4a)$$

$$h_b(\tau) = \left(\frac{2\tau_b - \tau}{\tau_b^2}\right) \exp\left(-\frac{\tau}{\tau_b}\right) \quad (4b)$$

and the coefficients $f_a = 0.75$, $f_b = 0.21$, and $f_c = 0.04$ quantify the relative contributions of the isotropic and anisotropic parts of the Raman response. In (4a) and (4b), $\tau_1$, $\tau_2$ and $\tau_b$ have values of 12, 32 and 96 fs, respectively. In our notation, they are normalized by the input pulse width $T_0$. Equation (1) is numerically solved by using split-step Fourier method [22] for a given input pulse having the form $u(0, \tau) = \sec h(\tau)$.

### B. Phase matching equation

Since the input pulse is launched in the normal dispersion domain it will not form a typical soliton, rather it reshapes and forms a rectangular shock wave with oscillating tail [21]. The resonant radiation will be emitted at the frequencies for which the propagation constant of the DW matches with the input pulse momentum. This condition can be found by equating the total phase of the pump ($\phi_0$) and radiation ($\phi_R$), which can simply be written as [2], [20], [22],

$$\phi_0 = \left[\beta(\omega_0) - \omega_0 / v_g(\omega_0) + k_{NL}\right] z \quad (5)$$

$$\phi_R = \left[\beta(\omega_R) - \omega_R / v_g(\omega_0)\right] z \quad (6)$$

where, $\omega_0$ and $\omega_R$ are the frequencies of pump and RR respectively. $k_{NL} = \gamma P_0 (1 + \Delta\omega / \omega_0)$, is nonlinear contribution in phase due to self-phase modulation and self-steepening. $\beta(\omega)$ is the propagation constant of the medium and $v_g = \partial_\omega \beta(\omega)^{-1}$ is the group velocity which is approximated to be same for pump and radiation. The phase matching condition ($\phi_0 = \phi_R$) leads to the equation of frequency detuning ($\Delta\omega = \omega_R - \omega_0$) between radiation and pump. Taylor series expansion of $\beta(\omega_R)$ around the pump wavelength ($\lambda_0 = 2\pi c \omega_0^{-1}$) results the following phase-matching expression,

$$\sum_{m=2}^{\infty} \frac{1}{m!} \beta_m \Delta\omega^m = k_{NL} \quad (7)$$

Equation (7) is a simple polynomial equation whose solutions predict the frequency of the RR. Neglecting the HOD terms for $m > 3$ (i.e $\beta_{m>3} = 0$) we can explicitly solve (7) and can have the following general solution,

$$\Delta\omega = -\frac{3\beta_2}{2\beta_3} \left[1 + \sqrt{1 + \frac{4\beta_3 \lambda_0}{3\pi c \beta_2^2} \gamma P_0}\right] \quad (8)$$

Here we neglect all higher order dispersion ($m > 3$) effects because of their insignificant contributions in frequency detuning. The traditional SPM-induced phase term is also neglected since the pulse is launched in ND domain; however we keep the shock term (second term under root) which may be significant for the pumping near zero dispersion wavelength. The negative sign in the right hand side of the equation clearly indicates that for $\beta_2 > 0$ and $\beta_3 > 0$, the radiation will be red-shifted. Note that the negative sign in the parenthesis is not considered here because of the negligible contribution of the shock term which eventually makes frequency detuning $\Delta\omega \approx 0$.

## IV. RESULTS AND DISCUSSION

In this section we demonstrate the experimental results which indicate the formation of solitonic RR and try to describe the underlining physics by comparing the experimental results with the numerical simulations. In Fig. 3 we have shown the evolution of the experimental spectra when a femtosecond laser pulse is launched at $\lambda_0 = 1064$ nm. We did our experiment for a fixed power at four different fiber lengths from 4 to10 m in the steps of 2 m. It is evident that the spectrum is initially broadened because of the self-phase modulation. However over the distance the broadening becomes asymmetric because of the Raman induced frequency red-shifting and eventually a significant radiation is observed around $\lambda_{RR} \approx 1140 nm$ at a distance of 10 m. We denote $\lambda_{RR}$ as the resonance wavelength for solitonic radiation. It is obvious that one can observe the RR at smaller distance by increasing the input power level. We have also done that experiment at high input power and observed the formation of RR even at 2 m of fiber length.

In the next phase of our work we try to corroborate the theoretical prediction with experimental results. In order to do so we numerically solve Equation (1) using the parameters which we have used in the real experiment. In Fig. 4 (a) we show the experimental spectrum (solid line) at the fiber output (z =10 m) when a *sech* pulse with temporal width of $T_0 \approx 275$ fs is launched at the wavelength of $\lambda_0 \approx 1064$ nm. The result is compared with numerically obtained spectrum (dotted) which is generated by solving the GNLSE as given in equation (1). In this numerical simulation we have taken the dispersion coefficient up to $8^{th}$ order. The simulated result agrees well with the experimental spectrum and both indicate a peak across the ZDW in the AD domain at the wavelength around $\lambda_{RR} \approx 1140 nm$. This red shifted peak is essentially a dispersive wave peak which is generated because of the phase matching between pump and DW. In order to ensure this we calculate the detuned wavelength using equation (8) by taking the fiber dispersion parameters $\beta_2 \approx 2.89$ ps$^2$/km and $\beta_3 \approx 0.065$ ps$^3$/km. The resonant wavelength ($\lambda_{RR}$) comes out to be 1144 nm which is very close to the experimentally observed peak. To explore the pulse dynamics we plot the temporal and spectral evolution of the input pulse over fiber length in Fig. 4(b) and 4(c), respectively. The temporal evolution in Fig 4 (b) shows a standard rectangular pulse reshaping with optical wave breaking that we always observe when we launch a pulse in ND domain. However the striking feature appears at a distance around $z$=8m when the spectrum overlaps with the ZDW. In temporal domain we find an interference pattern (as marked by the dotted curve in Fig. 4(b)). The XFROG analysis (Fig 4(d)) suggests, the temporal overlap between DW and the shock front of the leading edge of pump pulse produces interference fringe pattern in time domain. The spectral evolution of the pulse shows initial broadening due to self- phase modulation as expected when we launch the pulse in ND domain. However owing to the presence of Raman term the spectrum eventually becomes red shifted and overlaps with the zero dispersion point. This leads to the radiation across ZDW in AD domain. The radiation is indicated by dashed circle in Fig. 4(c). To capture the complete picture of the pulse dynamics we plot the cross-correlation frequency resolved optical gating (XFROG) spectrogram in Fig. 4(d). XFROG is a standard technique used to represent ultrashort pulses in frequency and time domain and is defined as the convolution $S = \left| \int u(\tau',\xi) u_{ref}(\tau-\tau') \exp(i\omega\tau') d\tau' \right|^2$, where $u_{ref}$ is the reference window function, generally taken as the input pulse. From the XFROG, the formation of solitonic-RR across ZDW is evident. The most interesting feature that is to be noted here is, the radiation is confined in both time and frequency (or wavelength) domain and essentially form a solitonic structure. The Raman red-shifted peak can also be identified in the same plot. We have investigated that even at long distances of the fiber the RR remains localized and maintains its solitonic nature. We also plot the group delay curve (dotted curve) in Fig. 4(d) and identify that the radiation wave is dragged by the leading shock front of the pump pulse. As a consequence of that there is a temporal overlap between RR and pump which produces interference fringe pattern in time domain which we have already seen in Fig. 4(b). Next we extend our study for larger pulse width. In Fig. 5 we show the experimental spectra for different fiber lengths. In this experiment we launch a picosecond pulse having the pulse width $T_0 \approx 5.5$ ps with fixed average input power of 830 mW. Here we increase the average input power level to excite the RR for approximately same fiber length (z=10 m). Since for picosecond pulse Raman effect is less predominant we observe less frequency red shifting here. However by increasing input power we eventually broaden the spectrum by means of SPM and at some point the spectrum overlaps the ZDW. The initial characteristics of the spectrum don't differ much with femtosecond results since in both the cases SPM plays the key role in broadening the spectrum.

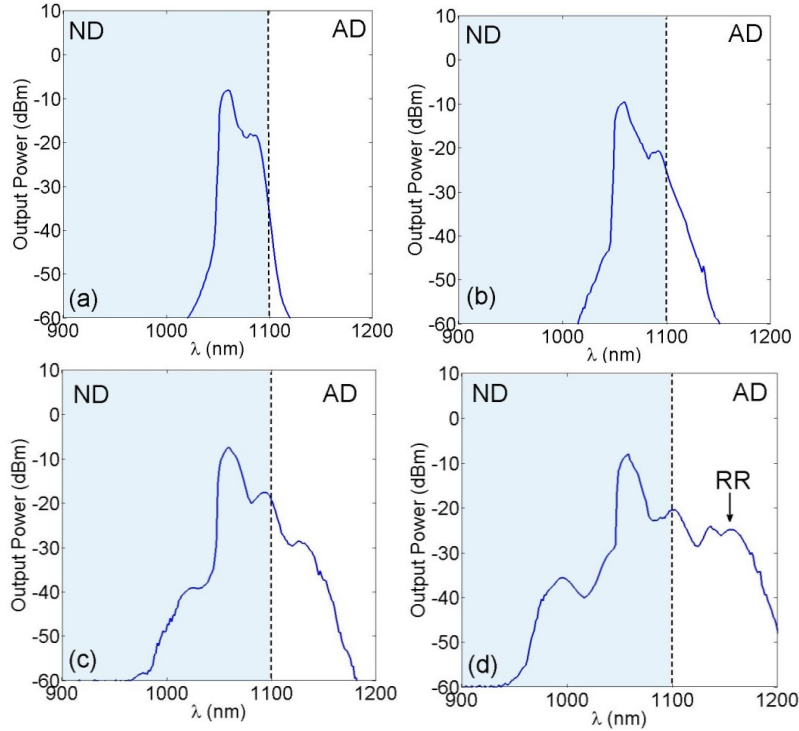

Fig. 3: The evolution of the experimental spectra is shown for the fiber length of (a) 4m, (b) 6m, (c) 8m and (d) 10m. The vertical dotted line separates the normal (shaded region) and anomalous dispersion. The input pulse width and average power are 275 fs and 440 mW respectively.

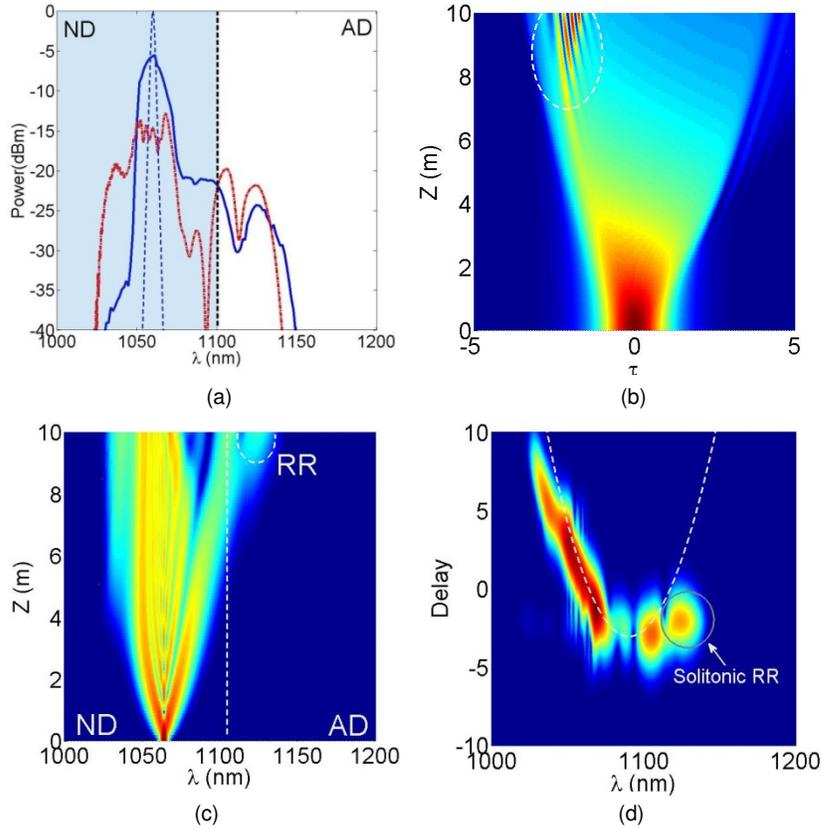

Fig. 4: (a) Experimental (solid blue line) and simulated (dotted red line) spectra at the output of 10 m fiber. The pulse width of the launched *sech* pulse is $T_0 \approx 275$ fs and the launching wavelength is $\lambda_0 = 1064$ nm. The average input power was 430 mW. The vertical dotted line differentiates the ND and AD regime. Dotted blue curve is the input spectrum. (b) Evolution of the pulse in time domain over the fiber length. The dotted region indicates the interference between shock front and solitonic radiation (c) Spectral evolution of the pulse over fiber length. Vertical dotted line indicates the zero dispersion wavelength. (d) Cross-correlation frequency resolved optical gating (XFROG) figure at fiber output ($z = 10\,m$) and the dotted white line represents the measured group-delay. The distinct solitonic-RRs are found around $\lambda_{RR} \approx 1140\,nm$.

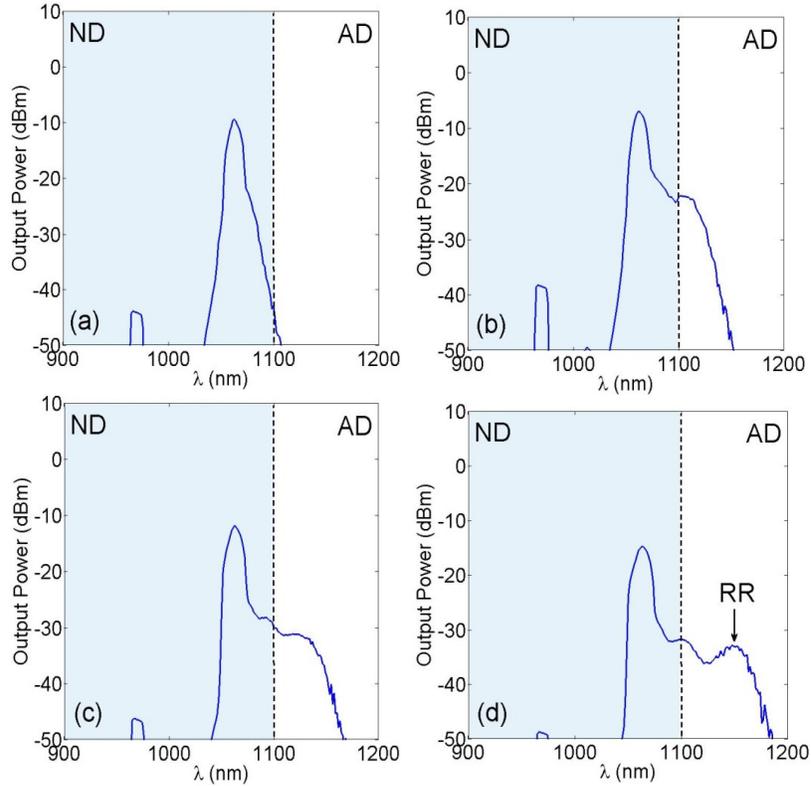

Fig. 5: The evolution of the experimental spectra is shown for the fiber length of (a) 4m, (b) 6m, (c) 8m and (d) 10m. The vertical dotted line separates the normal (shaded region) and anomalous dispersion regime. The input pulse width and average power is 5.5 ps and 830 mW respectively. The constant peak around 970 nm is the characteristic property of the laser that we used.

Interestingly the excitation wavelength of RR remains same for both the cases owing to the fact that radiation frequency is mainly governed by the dispersion profiles but not by the pulse widths.

In Fig. 6(a) we compare the experimental and numerical spectra which show the radiation peak around $\lambda_{RR} \approx 1140\,nm$. In Fig. 6(b) we plot the output pulse shape for an injected *sech* pulse in ND domain. The pulse shape significantly modified because of the generation of solitonic RR. The interference between the resonant RR and leading edge shock front of the pump pulse creates this pulse reshaping. Interestingly a similar observation was made in [24], [25] where the pulse dynamics were studied in ND regime. However the authors completely overlooked the spectral analysis where RR should appear. In plot 6(c) we have shown the usual spectral evolution over distance and observed that the RR appears at the distance of 10 m. Here also we perform the XFROG analysis to capture the entire picture of pulse dynamics. In Fig. 6(d) we plot the spectrogram which is obtained at z= 12 m. The spectrogram shows a similar pattern that we have already obtained in Fig. 4(d) except the weaker Raman Peak. The characteristic RR is distinctly observed in the figures along with the hint of the generation of Raman soliton. Raman soliton seeded by RR is a completely new feature that we have observed for long distances and describe in detail in the following section.

## V. GENERATION OF RR SEEDED RAMAN SOLITON

In this section we investigate theoretically the interplay between RR and IPRS. We simulate the NLSE for a relatively larger distance with a dispersion profile whose ZDW ($\lambda_{ZD} \approx 1110\,nm$) is slightly shifted to higher wavelength side. We intentionally shifted this ZDW to about 10 nm, which allow us to form a sharp radiation across $\lambda_{ZD}$ and one can identify IPRS and RR distinctly. In the previous case $\lambda_{ZD}(\approx 1100\,nm)$ was very close to the operating wavelength ($\lambda_0 \approx 1064\,nm$) and because of that it was difficult to identify the RR and Raman peak separately at a long distance since both are red-shifted. At this point it should be noted that in fabrication point of view this dispersion shifting is always possible by changing the drawing parameters of the PCF. Indeed, using Comsol Multiphysics software we have modeled the suspended core PCF for which $\lambda_{ZD} \approx 1110\,nm$ where the numeric values of the geometric parameters like, $d$, $\Lambda$ and $d_c$ turns out to be 7 $\mu m$, 8 $\mu m$ and 7 $\mu m$ respectively. Some fascinating dynamics are observed when we simulate the evolution of a femtosecond pulse spectrum over a long fiber length (z=20 m).

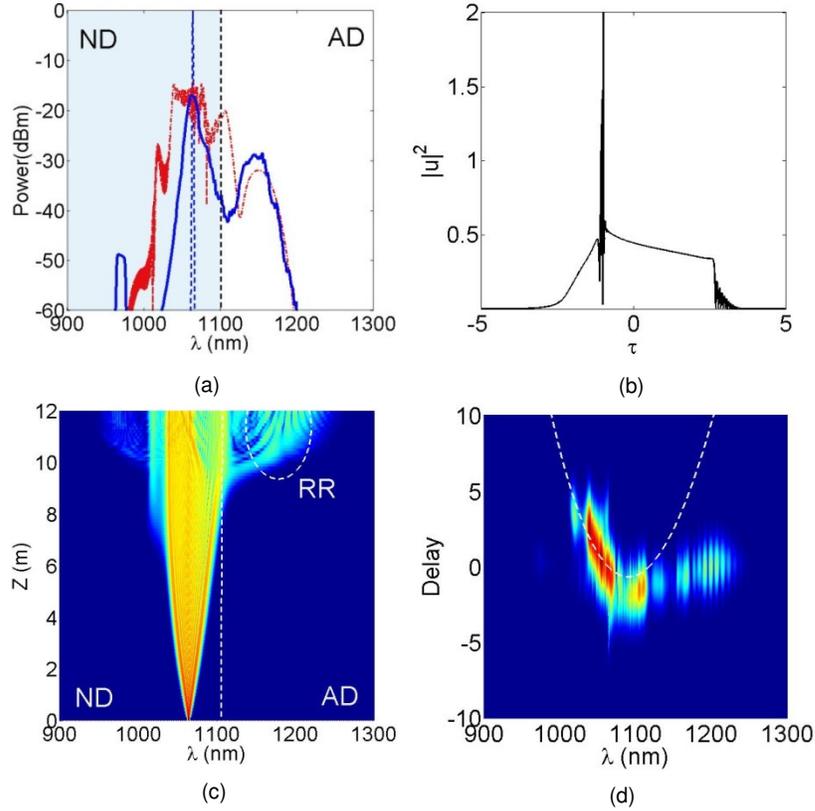

Fig. 6: (a) Experimental (solid blue line) and simulated (dotted red line) spectra at the output of 10 m fiber. The pulse width of the launched *sech* pulse is $T_0 \approx 5.5$ ps and the launching wavelength is $\lambda_0 = 1064$ nm. The average input power was 860 mW. The vertical dotted line differentiates the ND and AD regime. Dotted blue curve is the input spectrum. (b) Output pulse shape in time domain at the fiber length of $z = 12\,m$. (c) Spectral evolution of the pulse over fiber length. Vertical dotted line indicates the zero dispersion wavelength. (d) XFROG at fiber output ($z = 12\,m$) and the dotted white line represents the measured group-delay. The distinct solitonic-RRs are found around $\lambda_{RR} \approx 1140\,nm$.

As shown in Fig 7 (a), a pure RR appears first at the fiber length $z \approx 8\,m$ and then a part of the radiation gets red shifted under IPRS process and eventually forms Raman soliton. The formation of the Raman soliton is distinctly evident in the XFROG as shown in Fig. 7(b). Note that it is the *solitonic* RR that leads to the inception of Raman soliton in AD regime. When we extend our simulation to longer distance (z=50 m) more striking features are appeared. It is observed that instead of one soliton, the RR now leads to a bunch of Raman solitons marked as 1-4 in Fig. 7(c) and Fig. 7(d). The generation of this series of solitons is not typically because of the *soliton fission* [3], [22], [26] process where, under certain perturbation, a higher order soliton breaks into its fundamental components. On the contrary, here the individual solitons are seeded by the RR at different propagation distances. In Fig 7 (d) we have plotted the XFROG spectrogram where the formations of Raman solitons are evident. The intensity of the Raman soliton is stronger than the rest of the pulse. So the cross phase modulation (XPM) induced by the Raman soliton on the rest of the pulse is significant. The interactions of Raman solitons and the original pulse generate new frequencies (in the blue side) under the XPM process. This feature can only be observed in the XFROG spectrogram representation and in Fig. 7 (d) it is indicated by arrows.

## VI. CONCLUSION

In conclusion, we have experimentally verified the formation of red shifted RR in AD domain when we launch picosecond and femtosecond pulses in ND regime of PCF. We fabricated the PCF in such a way that the zero group velocity dispersion falls close to the operating wavelength which is at 1064 nm. In contrast to the conventional dispersive waves the phase matched RR is found to be solitonic in nature since it is confined both in time and frequency domain. The phase matching relationship between pump and radiation wave leads to the expression of the detuned frequency of the radiation. Using the coefficient of higher order dispersion we have calculated the resonance frequency by solving the phase matching equation and obtained close match with the experimental results. We performed a detailed numerical simulation for different pulse width regimes to understand the fascinating mechanism of the RR. Our theoretical study substantiate well with the experimental results. Finally we investigated the interplay between red shifted RR and IPRS

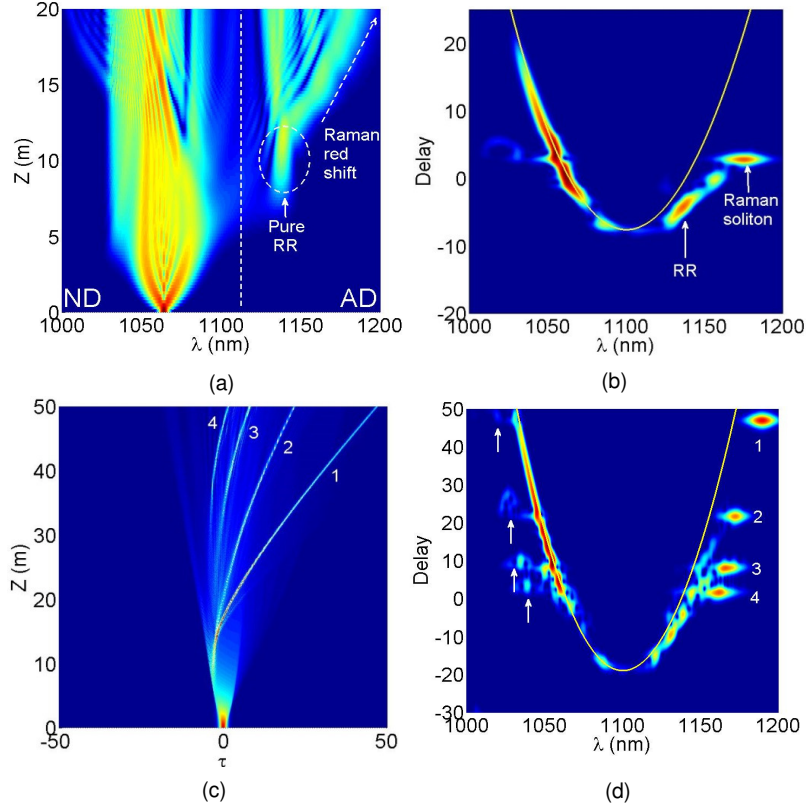

Fig. 7: (a) Spectral evolution of the pulse ($T_0 \approx 275$ fs) over fiber length. Vertical dotted line indicates the zero dispersion wavelength which is $\lambda_{ZD} \approx 1110$ nm with $\beta_2 \approx 4$ ps$^2$/km and $\beta_3 \approx 0.07$ ps$^3$/km. (b) XFROG at fiber output ($z=20 m$) and the solid line represents the measured group-delay. (c) Temporal evolution of the same pulse over a longer distance. The accelerated Raman solitons are indicated as 1-4. (d) XFROG at fiber output ($z=50 m$) where the formation of Raman solitons are shown distinctly. The new frequency components generated through XPM between Raman soliton and rest of the pulse are shown by arrows

for a fiber whose zero dispersion wavelength is slightly greater than our fabricated PCF. We observed unique features through XFROG analysis that first RR appears in the AD regime and then finally it leads to the generation of series of Raman solitons.

ACKNOWLEDGEMENT

The authors are grateful to the Director, CSIR-Central Glass and Ceramic Research Institute (CGCRI), Kolkata, for supporting and encouraging the work. Authors are indebted to Staff Members of FOPD, CGCRI for their sincere help.